\documentclass[aps,prd,showpacs,groupedaddress,floatfix,nofootinbib]{revtex4}
\usepackage{amsmath}
\usepackage{graphicx}
\usepackage{color}

\def\beq{\begin{eqnarray}}
\def\eeq{\end{eqnarray}}

\begin{document}

\title{Alternative representation for non--local operators and  path integrals}
\author{Paolo Amore\footnote{This paper is dedicated to the memory of Prof. Leopoldo Nieto Casas.}}
\email{paolo@ucol.mx}
\affiliation{Facultad de Ciencias, Universidad de Colima, \\
Bernal D\'{\i}az del Castillo 340, Colima, Colima, Mexico} 
\date{\today}

\begin{abstract}    
We derive an alternative representation for the relativistic non--local kinetic energy 
operator and we apply it to solve the relativistic Salpeter equation using the variational 
sinc collocation method. Our representation is {\sl analytical} and does not depend on an 
expansion in terms of local operators. We have used the relativistic harmonic oscillator 
problem to test our formula and we have found that arbitrarily precise results are obtained, 
simply increasing the number of grid points. More difficult problems have also been considered, 
observing in all cases the convergence of the numerical results.
Using these results we have also derived a new representation for the quantum mechanical 
Green's function and for the corresponding path integral.
We have tested this representation for a free particle in a box, recovering the exact result
after taking the proper limits, and we have also found that the application of the 
Feynman--Kac formula to our Green's function yields the correct ground state energy. 
Our path integral representation allows to treat hamiltonians containing non--local operators 
and it could provide to the community a new tool to deal with such class of problems.
\end{abstract}
\pacs{03.65.Ge,02.70.Jn,11.15.Tk}
\maketitle

%%%%%%%%%%%%%%%%%%%%%%%%%%%%%%%%%%%%%%%%%%%%%%%%%%

\section{Introduction}

The appearance of non--local operators in the relativistic extensions of the Schr\"odinger equation
poses a serious challenge both to analytical and numerical calculations. However, the inclusion of 
relativistic effects is crucial for example in the study of meson phenomenology, where the Bethe-Salpeter 
equation (BSE) provides the correct theoretical tool to describe relativistic bound states.
Replacing the kernel in the BSE with an instantaneous local potential one obtains a relativistic
Schr\"odinger equation, which is also known as ``spinless Salpeter equation'' (SSE). In such case
the hamiltonian operator is typically given by
\beq
\hat{H} = \sqrt{\hat{p}^2 + m^2} + V(\vec{r})  \ .
\label{eq1}
\eeq

From a technical point of view, the inclusion in the Hamiltonian of the relativistic kinetic energy 
operator, $\sqrt{\hat{p}^2 + m^2}$, complicates the solution of the problem because of its non--local 
nature. The great phenomenological relevance of the SSE has motivated in the past twenty years many 
efforts to solve this equation, either using analytical or numerical techniques. 
Early work on this  subject is contained for example in \cite{Durand84,Long84,Lucha92,Fulcher94, Maung94}.
The method described in \cite{Fulcher94}, which allows one to obtain a matrix representation of the 
non-local kinetic energy operator, has also been used recently \cite{Semay05} in conjunction with the
Lagrange mesh method. In this case, the method is particularly appealing since it does not require 
the evaluation of the matrix elements of the potential, but rather only the specification of the 
potential on the grid points. This feature is also shared by the so called sinc collocation methods 
(see for example \cite{Amore06a,Amore06b}), which could also be used straightforwardly together with the 
method of \cite{Fulcher94}: however we do not consider this possibility since we are rather interested in
developing a completely new approach, as it will soon become clear.

In a series of recent works, Lucha and collaborators have been able to obtain precise upper and lower 
bounds to the eigenvalues of the RSE, \cite{Lucha96,Lucha00,Lucha01,Lucha05a}: these bounds provide useful 
analytic or semi--analytic expressions and have been applied to a number of test problems (see for example 
\cite{Lucha05a}). Another approach has been followed in \cite{Lucha95}, where an effective 
hamiltonian of non-relativistic form, which includes relativistic effects by means of parameters 
depending on the momentum, was constructed. 

Finally, the SSE was also studied in the case of the relativistic harmonic oscillator (RHO) in three
recent papers, \cite{Znojil96,Mustafa99,Lucha05b}. In  \cite{Lucha05b}  the authors were able to obtain  recurrence
relations for the coefficients of the series defining the eigenfunctions. 
The reader can also find useful the detailed bibliographic information contained in that paper.

The main purpose of the present paper is to develop a completely new approach to the solution of the SSE:
we will derive an {\sl analytical} representation of the non--local kinetic energy operator and use it with
the Variational Sinc collocation Method (VSCM)~\cite{Amore06a} to obtain arbitrarily precise numerical results.
We will use the RHO of \cite{Lucha05b}, for which semi--analytical results are available, to test our approach.
The new representation that we have found for the relativistic kinetic energy operator has also been
generalized to the calculation of the quantum mechanical Green's function and has allowed us to obtain a 
new formula, which differs from the standard formula of path integrals.

The paper is organized as follows: in Section \ref{sec1} we describe the SSE for the RHO 
and obtain precise numerical solutions working in momentum space configuration (these results are then 
compared with the results of \cite{Lucha05b}); in Section \ref{sec2} we obtain an explicit analytical 
expression for the matrix elements of the non--local kinetic energy operator in terms of the so called 
``little sinc functions'' (LSF) recently studied in \cite{Amore06b} (these functions were first introduced by
Schwartz in \cite{Schwartz85} and later used by Baye in \cite{Baye95}); we use this representation in 
the VSCM working in coordinate space and reproduce the results previously obtained in momentum space; 
in Section \ref{sec3} we extend our method to obtain explicit analytical representation for the 
Green's function which holds for general potentials; finally in Section \ref{sec4} we draw our conclusions 
and set the direction for future work.

\section{The Relativistic Harmonic Oscillator}
\label{sec1}

The RHO problem corresponds to the case in which a spherical harmonic oscillator potential, 
$V(\vec{r}) = a r^2$, is used in the Hamiltonian of eq.~(\ref{eq1}). After a simple rescaling 
of the mass and of the energy the Schr\"odinger equation can be cast in the form 
\beq
\left[\sqrt{\hat{p}^2+\mu^2} + r^2 \right] |\psi\rangle = \varepsilon \ | \psi\rangle \ ,
\eeq
depending on a single parameter $\mu$~\footnote{We adopt the same notation used in \cite{Lucha05b}.}.
The advantage of considering this potential lies in the fact that the corresponding Schr\"odinger equation 
in momentum space representation is local and it can therefore be attacked with standard techniques. 
In this case we are left with the equation
\beq
\left[ - \Delta_p + \sqrt{p^2+\mu^2} \right]  \varPsi(p) = \varepsilon \ \varPsi(p) \ ,
\eeq
where $\Delta_p \equiv \frac{\partial^2}{\partial p_x^2} + \frac{\partial^2}{\partial p_y^2} + 
\frac{\partial^2}{\partial p_z^2}$. 

The authors of \cite{Lucha05b} focus their analysis on the $l=0$ solutions and obtain a recurrence relation
for the coefficients of the reduced radial wave function $y(p) = \sqrt{4\pi} p \ \varPsi(p)$. The same problem
has also been discussed by M. Trott in \cite{Trott}, using a basis of harmonic oscillator wave functions in momentum
space and numerically computing in this basis the matrix element $\langle i |\sqrt{ -\partial_x^2 + m^2} | j \rangle$.

We will use the VSCM of \cite{Amore06a, Amore06b} to obtain a numerical solution of the problem and 
then compare our solution with the results of \cite{Lucha05b}. The VSCM uses sinc functions (SF), defined 
on the real line, or ``little sinc functions'' (LSF)\cite{Amore06b}, a particular generalization of sinc 
functions defined on finite intervals, to solve the Schr\"odinger equation by a collocation technique. 
Since the details of this method are clearly explained in \cite{Amore06a,Amore06b}, we will here avoid mentioning all
the technical details focusing on giving a more qualitative picture.

SF and LSF are functions which are strongly peaked around a certain value and they fastly decay and oscillate, when
moving away from this value. Under certain conditions they can be chosen to be orthogonal: such set of orthogonal 
functions is obtained by performing a replica of one function at the points where this function 
vanishes, thus effectively introducing a grid. Unless otherwise specified, either by the Physics of the problem
or by convention, the spacing of the grid is {\sl arbitrary} and, if not carefully chosen, it strongly affects
the precision of the results. In a recent paper Amore et al. have used an arbitrary scale factor
as a variational parameter in the solution of the Schr\"odinger equation using a basis of simple harmonic oscillator
wave functions~\cite{Amore05a}: in that case it was shown that the optimal scale factor could be chosen to 
minimize the trace of the Hamiltonian matrix in the Hilbert subspace spanned by the $N$ elements of the basis. 
The same principle, which was inspired by the Principle of Minimal Sensitivity (PMS)~\cite{Ste81}, was then used
in \cite{Amore06a,Amore06b} using a sinc collocation technique. As mentioned in the Introduction the
sinc collocation has the great advantage of avoiding the evaluation of matrix elements of the potential, which
is simply ``collocated'' at the grid points; on the other hand the evaluation of the matrix elements of the
non--relativistic kinetic operator is also quite simple, since it involves matrices whose elements are obtained by 
collocating the derivative of a given sinc function at the grid points.
In this way one easily obtains a matrix representation for the Hamiltonian. 

For example in one dimension one has
\beq
H_{kl} = - \frac{\hbar^2}{2m} c_{kl}^{(2)} + \delta_{kl} V(k h)  \ .
\label{collocnr}
\eeq
where $c^{(2)}$ is the matrix obtained from the second derivative of the sinc functions. $h$ is the
grid spacing. In the following we will set $\hbar = 1$, so that $h$ will not be confused with the Planck constant.

The diagonalization of $H$ allows one to obtain the eigenvalues (energies) and the eigenvectors (wave functions)
of the problem (the number of these eigenvalues and eigenvectors being equal to the number of sinc functions
used). 

Before trying to deal with the RHO in coordinate space, where it is non--local, we wish to 
use the VSCM to find numerical solutions in momentum space, where it is local. In this case
machinery of VSCM applies straightforwardly and we are able to verify our claim of accuracy
of our method. To allow a comparison with \cite{Lucha05b} we have used $\mu = 30$ and calculated
the values of $\varepsilon-\mu$ for $\mu=30$ using grids with different $N$. The optimal region in
momentum space  has been obtained by applying the PMS condition to the problem. 
Our results show that the method converges quite rapidly; as a matter of fact, for $N = 50$ we obtain 
$\varepsilon -\mu = 0.386266042572445193517188444455$ which has all the digits correct and agrees
with the result shown in  Table 1 of \cite{Lucha05b}, although the latter contains only $6$ digits.

However we have also performed a more accurate test using the recurrence relations for the coefficients appearing
in the series for the wave functions given in \cite{Lucha05b}. We have extracted these coefficients 
by expanding around $p = 0$ the wave function obtained with our method and we have compared them 
with the results obtained using directly the recurrence relation of eq.(8) of \cite{Lucha05b}. 
In this way we have obtained an indipendent confirmation of the accuracy of our results.

The importance of the RHO lies in the unique possibility of having a complete control on the solutions 
of the problem, which can be calculated to any desired accuracy, despite retaining the non--local nature
of the kinetic energy operator. In more general problems, when the potential is not quadratic in the 
coordinates, one cannot recover a local Schr\"odinger equation by working in the momentum space 
configuration. In these cases one possibility is to resort to a non--relativistic expansion in powers
of $p/\mu$, which to lowest order provides the rest mass and the usual non--relativistic kinetic energy:
\beq
\sqrt{\hat{p}^2 + \mu^2} \approx \mu + \frac{\hat{p}^2}{2 \mu} + \dots 
\label{eq2}
\eeq
However the hamiltonian operator obtained in this way contains 
derivatives of higher order. In such cases it is still possible to apply the VSCM in coordinate space 
to solve these equations, by using the matrices corresponding to the higher order derivatives 
(which can be computed quite easily). 
Although the implementation of this procedure poses no problem, we wish to show that in certain cases it 
can provide unexpected results, such as results which converge to wrong solutions as the higher order
relativistic corrections are added.

We use once again the RHO as our ``guinea pig'' and work in momentum space configuration 
using a potential obtained by expanding the kinetic term up to a give order in $p^2/\mu^2$. The expression in
eq.~(\ref{eq2}), for example, would represent the potential obtained working to order $p^2/\mu^2$ and corresponds
to the usual non--relativistic harmonic oscillator. 
However, in the inclusion of the higher order terms we have to take into account that terms of order 
$(p^2/\mu^2)^{2 n}$, with $n$ integer, are negative, and therefore one always needs to work to order  
$(p^2/\mu^2)^{2 n+1}$ to have a spectrum bounded from below. We call $\tilde{V}_{2n+1}(p)$ the potential
in momentum space corresponding to the expansion to order $(p^2/\mu^2)^{2 n+1}$. 
For example to order $(p^2/\mu^2)^3$ we have
\beq
\tilde{V}_3(p) = \mu + \frac{p^2}{2 \mu } -\frac{p^4}{8 \mu ^3}+\frac{p^6}{16 \mu^5} \ .
\eeq

Once we substitute the non--local operator with its local expansion to a given odd order we can solve the
corresponding local Schr\"odinger equation and thus obtain the energies and the wave functions. 
We have calculated the values of $\varepsilon-\mu$ for the ground state of the potential 
using $N = 100$ and two different values of $\mu$ ($\mu = 30$ and $\mu=5$).  In the case corresponding to
$\mu =30$ we have observed that  the series converges to the exact result within $28$ digit precision 
for $\tilde{V}_{31}$, while in the case corresponding to $\mu = 5$ we have observed that the series
does not converge to the exact result, providing its best approximation for $\tilde{V}_9$.

We can give a simple explanation of this behavior: for $p >\mu$ the series in eq.~(\ref{eq2}) -- 
where $p$ is now a real number -- diverges and therefore the potential obtained using this series, 
$\tilde{V}_\infty(p) = \lim_{n\rightarrow\infty} \tilde{V}_{2n+1}(p)$, corresponds to the original 
potential with an infinite wall located at $p=\mu$. In the case $\mu = 30$ the wave function
is extremely small at $p = \mu$ and therefore the non--relativistic expansion is capable of providing 
highly (but not arbitrarily) accurate results; on the other hand, in the case $\mu =5$ the wave 
function is sizeable at $p=\mu$ and the non--relativistic expansion provides a very poor approximation.
The situation would clearly get worse as $\mu$ is taken to be smaller.

It should be stressed that in both cases the non--relativistic expansion does converge, although not 
to the exact result, but rather to a value which depends on the scale $\mu$.
I am not aware if this is a well-known fact in the literature.

\section{Matrix elements of $\sqrt{\hat{p}^2+\mu^2}$}
\label{sec2}

In this Section we want to describe a new way of calculating the 
matrix elements of the relativistic kinetic energy operator, which
avoids the problems of the local non--relativistic expansion described
in the previous Section.

As mentioned in the Introduction, there are approaches in the literature 
to calculate the matrix elements of this operator, although they are numerical.
For example, the method described in \cite{Fulcher94} consists of $4$ steps, i.e.
the computation of $(K^2)_{ij} = \langle i | \hat{p}^2+\mu^2 | j\rangle$,  
the diagonalization of $K^2$, the computation of the diagonal square root matrix 
and finally the computation of the square root matrix $K$. Because of the 
calculation of $K$ is numerical this procedure would not be profitable in a 
variational scheme, where the PMS condition is always analytical.

However, we will now show that it is possible to obtain {\sl once and for all}
an analytical representation of $\sqrt{\hat{p}^2+\mu^2}$ and indeed of any non--local
operator which is function of the momentum operator. Because our results will be 
analytical it will be possible to extend the VSCM to include relativistic non--local terms, 
in what we will call ``Relativistic variational collocation method'' (RVSCM).

Let us now present our results. We will work in the following with the Little Sinc 
Functions (LSF) of \cite{Amore06b}, although similar results can also be obtained 
with the usual sinc functions\footnote{As a matter of fact it is shown in \cite{Amore06b}
that for $N\rightarrow \infty$ and keeping $h$ fixed the LSFs reduce to the SFs.}.
The LSF have been obtained in \cite{Amore06b} using the orthonormal basis 
of the wave functions of a particle in a box with infinite walls located at 
$x=\pm L$:
\beq
\psi_n(x) = \frac{1}{\sqrt{L}} \ \sin \left(\frac{n\pi}{2L} (x+L)\right) \ .
\eeq
A LSF is simply obtained as
\beq
s_k(h,N,x) &=& \frac{2 L}{N}\sum_{n=1}^N \psi_n(x) \psi_n(y_k) \ ,
\eeq
where $y_k \equiv \frac{2 k L}{N} = k h$. The LSF should be regarded as an approximate
representation of a Dirac delta function. To simplify the notation we have introduced
the grid spacing $h \equiv 2L/N$. $k$ is an integer which takes the values 
 $k = - N/2+1, -N/2+2, \dots , N/2-1$, each corresponding to a different grid point.

As shown in \cite{Amore06b} an explicit expression for the LSF can be obtained for even values of $N$
in the form
\begin{widetext}
\beq
s_k(h,N,x) = \frac{1}{2 N}
\left\{ \frac{\sin \left( \left(1+\frac{1}{2N}\right) \ 
\frac{\pi}{h} (x-k h) \right)}{\sin \left( \frac{\pi}{2 N h} (x-k h) \right) }
-\frac{\cos\left(\left(1+\frac{1}{2N}\right) \ 
\frac{\pi}{h} (x+k h)\right)}{\cos \left(\frac{\pi}{2 N h} (x+k h)\right)} \right\} \ .
\label{sincls}
\eeq
\end{widetext}

After simple algebra one can also obtain the alternative expression
\begin{widetext}
\beq
s_k(h,N,x) = - \frac{1}{N} \sum_{n=-N}^{+N} (i)^{n+1} \ \sin \left( \frac{n \pi}{2} + \frac{k n \pi}{N}\right) \
e^{i \frac{n \pi x}{2 L}} 
\eeq
\end{widetext}
which is suited to  calculate the action of $\sqrt{\hat{p}^2+\mu^2}$ over a LSF:
\begin{widetext}
\beq
\sqrt{\hat{p}^2+\mu^2} \  s_k(h,N,x) = - 
\frac{1}{N} \sum_{n=-N}^{+N} (i)^{n+1} \ \sin \left( \frac{n \pi}{2} + \frac{k n \pi}{N}\right) \
\sqrt{\left(\frac{n \pi}{2 L}\right)^2 +\mu^2} \ e^{i \frac{n \pi x}{2 L}} \ .
\eeq
\end{widetext}
Upon collocation on the grid, i.e. after setting $x \rightarrow x_j$, we have the matrix element
\begin{widetext}
\beq
K_{kj} =  \sqrt{\hat{p}^2+\mu^2} \  s_k(h,N,x_j) = - 
\frac{1}{N} \sum_{n=-N}^{+N} (i)^{n+1} \ \sin \left( \frac{n \pi}{2} + \frac{k n \pi}{N}\right) \
\sqrt{\left(\frac{n \pi}{2 L}\right)^2 +\mu^2} \ e^{i \frac{j n \pi}{N}} \ .
\label{Kmat}
\eeq
\end{widetext}
Notice that, despite its appearance, the expression above is real, as it would be evident 
expressing it in terms of trigonometric functions; however, we prefer to leave it in terms of plane waves.

To check that eq.~(\ref{Kmat}) is indeed the correct matrix element of the non--local kinetic
energy operator we can calculate the matrix product $(K^2)_{kl} = \sum_{j} K_{kj} K_{jl}$ 
\beq
(K^2)_{kl} &=&  \sum_{j} K_{kj} K_{jl} = 
\frac{1}{N^2} \sum_{j=-N/2+1}^{N/2-1} \sum_{n_1=-N}^{+N} \sum_{n_2=-N}^{+N} 
(i)^{n_1+n_2+2} \ \sin \left( \frac{n_1 \pi}{2} + \frac{k n_1 \pi}{N}\right) \
\ \sin \left( \frac{n_2 \pi}{2} + \frac{j n_2 \pi}{N}\right) \nonumber \\
&&\sqrt{\left(\frac{n_1 \pi}{2 L}\right)^2 +\mu^2} \
\sqrt{\left(\frac{n_2 \pi}{2 L}\right)^2 +\mu^2} \  e^{i \frac{(j n_1 +l n_2) \pi}{N}}
\label{ksquare}
\eeq
and  compare it with the matrix $(K^2)_{kl} =(\hat{p}^2+\mu^2)_{kl}$, calculated either using 
the very same eq.~(\ref{Kmat}) with $\sqrt{\left(\frac{n \pi}{2 L}\right)^2 +\mu^2}
\rightarrow \left(\frac{n \pi}{2 L}\right)^2 +\mu^2$ or using the matrix for the second 
derivative to represent $\hat{p}^2$ (see \cite{Amore06b}).

We can define
\beq
{\cal C}_{n_1 n_2} \equiv 
\sum_{j=-N/2+1}^{N/2-1} e^{i \frac{j n_1  \pi}{N}}  \sin \left( \frac{n_2 \pi}{2} + \frac{j n_2 \pi}{N}\right) 
\eeq
which for $n_2 =\pm n_1$ takes the values
\beq
{\cal C}_{n_1 n_1} = -{\cal C}_{n_1, -n_1} = -\frac{N}{2} (-i)^{n1+1} \ .
\eeq

When ${\cal C}_{n1n2}$ is used inside eq.~(\ref{ksquare}) it can be seen that only the terms corresponding to
$n_2 = \pm n_1$ contribute, whereas the remaining terms cancel out. In this case we obtain
\beq
(K^2)_{kl} &=&  \frac{1}{N^2} \sum_{n_1=-N}^{+N} \sum_{n_2=-N}^{+N} 
(i)^{n_1+n_2+2} \ \sin \left( \frac{n_1 \pi}{2} + \frac{k n_1 \pi}{N}\right) \
\sqrt{\left(\frac{n_1 \pi}{2 L}\right)^2 +\mu^2} \
\sqrt{\left(\frac{n_2 \pi}{2 L}\right)^2 +\mu^2} \  e^{i \frac{l n_2 \pi}{N}} \nonumber \\
&& \left( -\frac{N}{2} (-i)^{n1+1} \right)  \left(\delta_{n_1,n_2}-\delta_{n_1,-n_2} \right) \nonumber \\
&=& -\frac{1}{N} \sum_{n=-N}^{+N} (i)^{n+1} \ \sin \left( \frac{n \pi}{2} + \frac{k n \pi}{N}\right) \
\left[\left(\frac{n \pi}{2 L}\right)^2 +\mu^2\right] \ e^{i \frac{j n \pi}{N}} \ .
\eeq

This completes our proof, since we have precisely obtained the expression that we would have
reached if we had used eq.~(\ref{Kmat}) directly with the squared operator. We have also verified 
numerically this result and confirmed its validity. 

As done for the non--relativistic Schr\"odinger equation we can obtain the Hamiltonian matrix 
\beq
H_{kl} = K_{kl} + \delta_{kl} V(k h)  \ .
\label{collocrel}
\eeq

For a fixed $N$ the matrix elements $H_{kl}$ depend upon the arbitrary scale $L$: as discussed 
in \cite{Amore06a,Amore06b}, it is possible to obtain an optimal value for $L$ by applying
the principle of minimal sensitivity (PMS) to the trace of the $N\times N$ hamiltonian matrix 
${\cal T}_N(L) = Tr\left[ \mathbf{H}_{N\times N}\right]$:
\beq
\frac{d}{dL} {\cal T}_N (L) = 0.
\label{pmseq}
\eeq

Physically one can justify this condition by observing that the trace of the hamiltonian is an invariant
under unitary transformations and therefore independent of $L$. On the other hand the trace of the truncated
matrix does depend on $L$: one can therefore apply the PMS condition (\ref{pmseq}) to minimize such dependence
and thus obtain the scale where the problem is less sensitive to changes in $L$  (normally this equation 
provides a single solution, although in the case of multiple solutions one would choose the solution corresponding 
to a flatter curve).
Since (\ref{pmseq})  is an algebraic equation, the computational cost needed in solving it is quite limited.
The scale $L$ obtained in this way is then used inside eq.~(\ref{collocrel}) and the hamiltonian matrix is 
diagonalized, thus providing the $N$ lowest eigenvalues and eigenfunctions. The error corresponding to 
choosing $L$ according to the PMS is observed to be nearly optimal, leading to results which converge more rapidly 
as $N$ increases.

\begin{figure}
\begin{center}
\includegraphics[width=9cm]{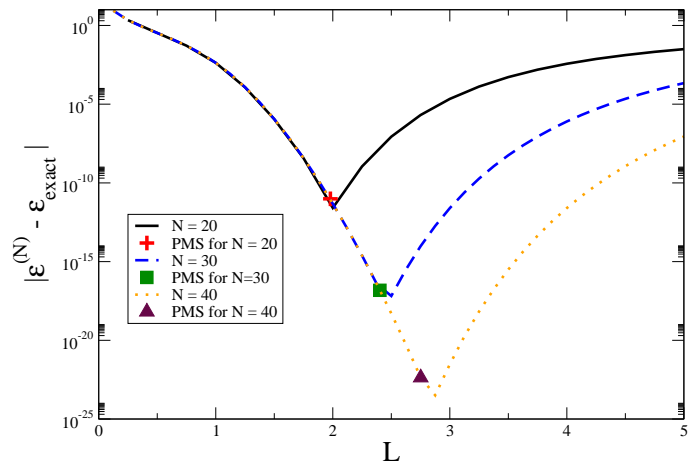}
\caption{$|\varepsilon^{(N)} - \varepsilon_{exact}|$ for the RHO as a function of $L$ for $\mu = 30$. 
The results of the PMS are also displayed. (color online)}
\bigskip
\label{FIG_PMS}
\end{center}
\end{figure}

We have used eq.~(\ref{collocrel}) in the case of the RHO with $\mu = 30$ and $\mu =5$, i.e. the case
which was previously studied in the momentum representation. The numerical results that we have obtained
in this nonlocal representation display a rate of convergence similar to the one observed in the local 
representation. In Fig.~\ref{FIG_PMS} we have plotted the difference 
$|\varepsilon^{(N)} - \varepsilon_{exact}|$ as a function of the scale $L$ and compared the results with the
predictions obtained using the PMS condition eq.~(\ref{pmseq}). Here $\varepsilon^{(N)}$ is the energy for the 
ground state obtained using a given value of $N$, whereas $\varepsilon_{exact}$ is approximated with the
energy obtained using $N = 100$.

In the case $\mu =5$ where the non-relativistic expansion provides a quite poor approximation
with just a four digit precision, the result obtained in the non--local representation has $12$ correct digits 
for $N=50$: $(\varepsilon - \mu) = 0.91531941208$.

Although the results obtained studying the RHO are sufficient in our opinion to prove the efficiency and simplicity
of the method that we are proposing in this paper, we wish to consider few examples which can better illustrate 
the power of our method.

As a first example we have chosen  the  potential
\beq
V(x) = \sqrt{a^2 +x^2}
\label{potential}
\eeq
and the corresponding SSE (with $x \in (0,\infty)$):
\beq
\left[\sqrt{\hat{p}^2 + \mu^2} + \sqrt{a^2+x^2} \right] \ \psi(x) = \varepsilon \ \psi(x) \ .
\eeq
Notice that for $a \gg 1$ and $\mu \gg 1$ one recovers the standard harmonic oscillator (SHO).
We have applied our method using  $a=\mu=1$ and we have obtained $(\varepsilon - \mu) =  2.075870921$, using 
$N = 100$, where all the displayed digits are correct. 

As a second example that we wish to consider the SSE (with $x \in (0,\infty)$):
\beq
\left[\sqrt{\hat{p}^4 + \hat{p}^2 + \mu^2} + \sqrt{a^2+x^2} \right] \ \psi(x) = \varepsilon \ \psi(x) \ .
\label{potential2}
\eeq
Also in this case we have  assumed $\mu=a=1$ and we have obtained $(\varepsilon - \mu) = 2.724199599$, using 
$N = 100$, where all the displayed digits are correct. 

As a final example we have considered the equation
\beq
\left\{\mu \ \log \left[1 + \frac{\hat{p}^2}{2 \mu^2} \right] + V(x) \right\} \psi(x) = E \ \psi(x) \ .
\eeq
with $V(x)=x^2/2$ and $x \in (-\infty,+\infty)$. Notice that the equation obtained expanding to leading order
in powers of  $p/\mu$ is the standard Schr\"odinger equation for the harmonic oscillator.
Despite the rather ugly form of the ``kinetic'' operator in the present case, the application of our method 
is straightforward and an expression for the matrix elements of the Hamiltonian is easily obtained:
\beq
H_{kj} =  - \frac{\mu}{N} \sum_{n=-N}^{+N} (i)^{n+1} \ \sin \left( \frac{n \pi}{2} + \frac{k n \pi}{N}\right) \
\log \left\{1+ \frac{n^2 \pi^2}{8 \mu^2 L^2}\right\} \ e^{i \frac{j n \pi}{N}}  + \delta_{kj} \ V(x_k) \ .
\eeq

We have solved this problem assuming $\mu=10$ and we have obtained the ground state energy 
$\varepsilon  =20.04909725  $  using $N=30$ (all the displayed digits are correct).  

Although we did not have any physical model in mind when we considered the last two examples, we are aware that
modifications of the standard Schr\"odinger equation (not necessarily of this kind) appear in several areas of 
research: for example, the introduction of a minimal length uncertainty relation
naturally leads to modified Schr\"odinger equations (see for example \cite{Kempf95}).

\section{Green's functions}
\label{sec3}

In this Section we want to show that the method that we have developed in the previous Section 
can also be applied to the calculation of matrix elements of other non--local operators. 
We will see that SF and LSF are a  powerful tool, and actually they have already been used 
earlier in a different context  in applications to Quantum Field Theory (QFT) (see for example \cite{Gura00}).

To start we consider an operator $\hat{O} \equiv f(\hat{p})$: in such case, the results
that we have obtained in the previous Section apply straightforwardly and one obtains
\begin{widetext}
\beq
O_{kj} = \frac{1}{N} \sum_{n=-N}^{+N} (i)^{n+1} \ \sin \left( \frac{n \pi}{2} + \frac{k n \pi}{N}\right) \
f\left(i \frac{n \pi}{2 L}\right) \  \ e^{i \frac{j n \pi}{N}} \ .
\eeq
\end{widetext}

One example of application of this formula is the calculation of the Green's function 
for a free particle in a box, $|x|<L$, which is given by (see for example \cite{Feynman,Schulman})
\beq
G^{(0)}(y,0;x,t) &=& \langle x | e^{-i \frac{\hat{p}^2}{2\mu} t} | y \rangle \nonumber \ .
\eeq

The notation $|x\rangle$ indicates a state localized at a point $x$:  in our formalism this is simply 
represented by a SF or a LSF with a peak at this point. Notice however that SF and LSF are not normalized
to one, see \cite{Amore06a,Amore06b}, which means that we have an extra factor $1/\sqrt{h}
\equiv \sqrt{N/2 L}$ for each SF or LSF, where $h$ is the spacing of the grid.

Therefore we can write:
\beq
G^{(0)}(x_j,0;x_k,t) &=& - \frac{1}{N h} \sum_{n=-N}^{+N} (i)^{n+1} \ 
\sin \left( \frac{n \pi}{2} + \frac{k n \pi}{N}\right) \
e^{-i t \left[\frac{1}{2\mu} \left(\frac{n \pi}{2 L}\right)^2 \right]}
\ e^{i \frac{j n \pi}{N}} \ ,
\eeq
$x_k$ and $x_j$ being points on the grid. In the limit of an infinitely dense grid one can 
switch from the discrete indices $k,j$ to the continuum indices $x,y$:
\beq
G^{(0)}(y,0;x,t) &=& -\frac{1}{2 L} \ \sum_{n=-N}^{+N} i^{n+1} e^{\frac{i n \pi  x}{2 L}-
\frac{i n^2 \pi ^2 t}{8 L^2 \mu }} \ \sin \left(\frac{\pi  y n}{2 L}+\frac{\pi  n}{2}\right)
\eeq

Notice that in the case of a free particle the exact result can be calculated with the standard path integral and
reads
\beq
G^{(0)}(0,0;x,t) = \sqrt{\frac{\mu}{2 \pi i t}} \ e^{i \frac{\mu x^2}{2 t}} \ .
\label{freeexact}
\eeq

In our formalism we have
\beq
G^{(0)}(0,0;x,t) &=&  -\frac{1}{2 L} \ \sum_{n=-N}^{+N} i^{n+1} e^{\frac{i n \pi  x}{2 L}-
\frac{i n^2 \pi ^2 t}{8 L^2 \mu }} \ \sin \left(\frac{\pi  n}{2}\right) \\
&=&\frac{1}{2 L}  \sum_{n=-N/2}^{+N/2} \ e^{-\frac{i \pi ^2 t n^2}{2 L^2 \mu }-\frac{i \pi ^2 t n}{2 L^2 \mu \
}+\frac{i \pi  x n}{L}-\frac{i \pi ^2 t}{8 L^2 \mu }+\frac{i \pi  x}{2 L}}
\eeq

For $N\gg 1$ we can approximate this sum with an integral and thus obtain
which reads
\beq
G^{(0)}(0,0;x,t) &=& \frac{1}{2} \ \sqrt{\frac{\mu}{2 \pi i t}} \ e^{i \frac{\mu x^2}{2 t}}
\left[\text{Erf}\left(\frac{\left(\frac{1}{4}+\frac{i}{4}\right) ((N+1) \pi  t-2 L \
\mu  x)}{L \sqrt{\mu t}}\right)+\text{Erf}\left(\frac{\left(\frac{1}{4}+\frac{i}{4}\right) \
((N-1) \pi  t+2 L \mu  x)}{L \sqrt{\mu t}}\right)\right] \ .
\eeq
In the limit $N \rightarrow \infty$ we can approximate each error function to one and therefore obtain
the exact expression (\ref{freeexact}).

To further test this formula we can also  use the Feynman-Kac formula to extract the ground state energy
\beq
E_0 = - \lim_{\tau\rightarrow\infty} \frac{1}{\tau}  \log \int dx \ G(x,-i \tau,x,0) \ .
\label{feyn}
\eeq

In our case the integral appearing in (\ref{feyn}) becomes a sum over all the grid points and 
therefore reads
\beq
E_0 = - \lim_{\tau\rightarrow\infty} \frac{1}{\tau} \log   \left\{
\frac{1}{N h} \sum_{k=-N/2+1}^{N/2-1}\sum_{n=-N}^{+N} (i)^{n+1} \ \sin \left( \frac{n \pi}{2} 
+ \frac{k n \pi}{N}\right) \ e^{- \tau \left[\frac{1}{2\mu} \left(\frac{n \pi}{2 L}\right)^2  \right]}
\ e^{i \frac{k n \pi}{N}} \right\}  .
\eeq

Because we are taking the limit $\tau \rightarrow \infty$, the exponential factor in this expression 
will be quite small unless $n=\pm 1$ (the term $n=0$ vanishes). For this reason we can 
approximate
\beq
E_0 &=& - \lim_{\tau\rightarrow\infty} \frac{1}{\tau} \log   \left\{
\frac{1}{N h} \sum_{k=-N/2+1}^{N/2-1} 
\left[2 e^{-\frac{\pi ^2 \tau }{8 \mu L^2}} \cos ^2\left(\frac{k \pi }{N}\right)\right] \right\} \nonumber \\
&=& - \lim_{\tau\rightarrow\infty} \frac{1}{\tau} \log   \left\{ e^{-\frac{\pi ^2 \tau }{8 \mu L^2}}
\right\} = \frac{\pi ^2}{8 \mu L^2}
\eeq
which is indeed the exact ground state of the particle in the box.

We can now consider the more general case in which the Hamiltonian contains a potential. In this case the 
formula of Section \ref{sec3} cannot be applied directly because $\hat{p}$ and $\hat{x}$ do not commute. 
However we can use the Trotter product formula to write~\cite{Feynman,Schulman}
\beq
G(y,0;x,t) &=& \lim_{\bar{N} \rightarrow \infty}  
\langle x | \left(e^{-i \frac{\hat{p}^2}{2\mu \bar{N}} t} 
e^{-i \frac{V}{ \bar{N}} t} \right)^{\bar{N}} | y \rangle \ .
\eeq

We can use the completeness of the coordinate states, $|x\rangle$, to write
\beq
G(y,0;x,t) &=& \lim_{\bar{N} \rightarrow \infty} \sum_{r_1,r_2,\dots, r_{\bar{N}}}
\langle x | \left(e^{-i \frac{\hat{p}^2}{2\mu \bar{N}} t} e^{-i \frac{V}{ \bar{N}} t} \right)| 
x_{r_1} \rangle \langle x_{r_1} | \left(e^{-i \frac{\hat{p}^2}{2\mu \bar{N}} t} 
e^{-i \frac{V}{ \bar{N}} t} \right)| x_{r_2} \rangle
\dots \langle x_{r_{\bar{N}}} | \left(e^{-i \frac{\hat{p}^2}{2\mu \bar{N}} t} 
e^{-i \frac{V}{ \bar{N}} t} \right)| y \rangle \ ,
\eeq
where the indices $r_i$ span all the lattice. Notice the factors $t/\bar{N}$, which correspond
to having a ``time--slicing'', as usual in path integration; at each intermediate time each point on the 
grid can be reached.

We define 
\beq
{\cal V}_{\bar{N}}(x_i,t) \equiv h \ \langle x_i | e^{-i \frac{V(x)}{ \bar{N}} t} | x_{j} \rangle  
= e^{-i \frac{V(x_i)}{ \bar{N}} t} \delta_{ij} \ ,
\eeq
which is diagonal in the coordinates. We can now write the  compact expression
\beq
G(y,0;x,t) &=& \lim_{\bar{N} \rightarrow \infty} \frac{1}{h^{\bar{N}}} \ 
\sum_{r_1,r_2,\dots, r_{\bar{N}}}
G^{(0)}(x_{r_1},t-\varepsilon;x,t) \ {\cal V}_{\bar{N}}(x_{r_1},t) \ 
G^{(0)}(x_{r_2},t-2\varepsilon;x_{r_1},t-\varepsilon) \nonumber \\ 
&&
{\cal V}_{\bar{N}}(x_{r_2},t)  \ \dots \ {\cal V}_{\bar{N}}(x_{r_{\bar{N}}},t) \ 
G^{(0)}(y,0; x_{r_{\bar{N}}},\varepsilon)  \ , 
\label{GF}
\eeq
where $\varepsilon \equiv t/\bar{N}$.
This expression  should be compared with the standard path integral expression\cite{Feynman,Schulman}
\beq
G(x,t;y,0) &=&  \lim_{N \rightarrow \infty} \ \int dx_1 \dots dx_{N-1} \ 
\left( \frac{\mu}{2\pi i \epsilon}  \right)^{N/2} \  e^{ i \epsilon \sum_{j=0}^{N-1} \left[
\frac{\mu}{2} \left( \frac{x_{j+1}-x_j}{\epsilon} \right)^2  -V(x_j) \right] } \ ,
\eeq
where $\epsilon \equiv t/N$.

Our equation (\ref{GF}) can be also compared with the exact expressions for the $N$-fold time sliced 
spacetime propagators which Crandall has obtained in eqs.(2.9) and (2.12) of \cite{Crandall93} using the standard representation.

Our result of eq.~(\ref{GF}) provides {\sl a new representation of the path integral for quantum mechanical 
problems} with a clear physical interpretation: the propagation of a particle sitting at $y$ at time $t=0$ 
and reaching $x$ at time $t$ occurs moving at each time interval, $\varepsilon$, from a point of the grid 
to another one in all possible ways (remember the sums over the grid points). $G^{(0)}(x_i,t;x_{j},t+\varepsilon)$ 
represents the probability of going from a point $x_i$ to another point $x_j$ on the grid in a time interval 
$\varepsilon$; at this point an interaction take place, through the the potential term ${\cal V}$. 
We stress that our representation corresponds to a different way of discretizing space which also
allows to deal with Hamiltonians containing non--local operators (the SSE is one example) and could be an useful tool
for problems which cannot be easily treated with the conventional formalism.

\section{Conclusions}
\label{sec4}

In this paper we have derived an analytical expression for the non--local relativistic kinetic 
energy operator which appears in the Salpeter equation. This representation is exact in the 
limit of an arbitrarily fine grid and we have used it to solve the Salpeter equation for the 
Relativistic Harmonic Oscillator, where semi--analytical results are available. 
We have found that our representation can be used together with the Variational Sinc Collocation 
Method (VSCM) to provide arbitrarily precise results, with a strong rate of convergence. 
The most important result of this paper is the new representation for the quantum mechanical 
Green's function, which requires the evaluation of matrix elements of non--local operators. 
We have provided a general formula and we have explicitly tested it in the case of a free particle 
in a box, recovering the exact result. 
Our new representation of the path integral can be applied also to problems in which the Hamiltonian
contains non--local operators (which would be the case of the Salpeter equation) and it  
is suitable both for numerical and  analytical calculations, in the cases in which the limits  
$N,\bar{N} \rightarrow \infty$ can be calculated (as for a free particle in a box).

Given the importance of path integrals in many areas of Physics, we feel that the results contained in this
paper could have a large number of applications. Finally, we wish to mention that 
it would be worth exploring the possibility to use the PMS in a numerical scheme to optimize convergence 
to the exact result for finite values of $N$ and $\bar{N}$. It also remains to explore the possibility
to apply our formalism to Quantum Field Theory (QFT), which we plan to address in future works.


\begin{thebibliography}{}
\bibitem{Durand84} L.J.Nickisch, L.Durand and B.Durand, Phys. Rev. D{\bf 30}, 660 (1984)
\bibitem{Long84} C. Long, Phys. Rev. D {\bf 30}, 1970 (1984)
\bibitem{Lucha92} W. Lucha, H. Rupprecht  and F.F. Sch\"oberl, Phys. Rev. D {\bf 45}, 1233-1239 (1992)
\bibitem{Fulcher94} L.P. Fulcher, Phys. Rev. D {\bf 50}, 447 (1994)
\bibitem{Maung94} J.W. Norbury, K. Maung Maung and D. E. Kahana, Phys. Rev. D {\bf 50}, 3609 (1994) 
\bibitem{Semay05} C. Semay, D. Baye, M. Hesse and B. Silvestre-Brac, Phys. Rev.E {\bf 64}, 016703 (2005)
\bibitem{Amore06a} P. Amore, A variational sinc collocation method for strong coupling problems, Journal
of Physics {\bf A} 39, L349-L355 (2006)
\bibitem{Amore06b} P. Amore, M. Cervantes and F.M. Fern\'andez, ArXiv:[quant-ph/0608069] (2006)
\bibitem{Lucha96} W. Lucha and F.F. Sch\"oberl, Phys. Rev. A {\bf 54}, 3790-3794 (1996)
\bibitem{Lucha00} W. Lucha and F.F. Sch\"oberl, J. Math. Phys.41, 1778 (2000)
\bibitem{Lucha01} R. Hall, W. Lucha and F.F. Sch\"oberl, J. Phys. A {\bf 34}, 5059-5063 (2001)
\bibitem{Lucha05a} R. Hall and W. Lucha, J. Phys. A {\bf 38}, 7977-8002 (2005)
\bibitem{Lucha95} W. Lucha and F.F. Sch\"oberl, Phys. Rev. A {\bf  51}, 4419- 4426 (1995)
\bibitem{Znojil96} M.Znojil, J. of Phys. A {\bf 29}, 2905 (1996)
\bibitem{Mustafa99} O. Mustafa and M. Odeh,  J. of Phys. A {\bf 32}, 6653 (1999)
\bibitem{Lucha05b} Li, Liu, W. Lucha, Ma and F.F. Sch\"oberl, J. Math. Phys.46, 103514 (2005) 
\bibitem{Schwartz85} C. Schwartz, Journal of Mathematical
Physics 26, 411-415 (1985)
\bibitem{Baye95} D. Baye, Journal of Physics {\bf B} 28, 4399-4412 (1995)
\bibitem{Trott} M. Trott, The Mathematica guidebook for symbolics, Springer
\bibitem{Amore05a} P. Amore, A. Aranda, F.M. Fern\'andez and H.F. Jones, Physics Letters A 340, 87-93 (2005)
\bibitem{Ste81}  P. M. Stevenson, Phys. Rev. D {\bf 23}, 2916 (1981).
\bibitem{Kempf95} A. Kempf, G.Mangano and R.B. Mann, Phys. Rev. D {\bf 52} 1108 (1995)
\bibitem{Gura00} R. Easther, G.Guralnik and S.Hahn, Phys. Rev. D {\bf 61}, 125001 (2000)
\bibitem{Feynman} R.P.Feynman and A.R.Hibbs, Quantum mechanics and path integrals, McGraw-Hill, New York (1965)
\bibitem{Schulman} L.S. Schulman, Techniques and applications of Path Integration, John Wiley and Sons, New York (1981)
\bibitem{Crandall93} R.E. Crandall, J. Phys. A {\bf 26}, 3627-3648 (1993)
\end{thebibliography}
\end{document}